\documentclass[11pt]{article}
\usepackage[dvips]{graphics,color}
\usepackage{moriond,epsfig}

\def\be{\begin{equation}}
\def\ee{\end{equation}}
\def\bea{\begin{eqnarray}}
\def\eea{\end{eqnarray}}

\begin{document}
\vspace*{4cm}
\title{The recent results of solar neutrino measuerments in Super-Kamiokande}

\author{Yusuke Koshio\\(for Super-Kamiokande collaboration)}

\address{Kamioka observatory, ICRR, Univ. of Tokyo\\
Higashi-Mozumi Kamioka-cho Yoshiki-gun Gifu pref. Japan}

\maketitle\abstracts{
The results of solar neutrino data from the first phase of
Super-Kamiokande are presented. Super-Kamiokande can measure not only the
solar neutrino flux but also its energy spectrum and its time variations
such as day vs. night and seasonal differences.  This information can
severely restrict parameters of solar neutrino oscillation. From the
combination of several experiments' results with those of Super-K, the
Large Mixing Angle solution is uniquely allowed at the 98.1\% confidence
level; this global solar neutrino oscillation analysis is presented.
The current status of the second phase of Super-Kamiokande is presented.}

\section{Introduction}
\subsection{Solar Neutrinos}
 The origin of the energy in the sun is the following nuclear fusion reaction
which generates two neutrinos,
\begin{equation}
  4p \to \alpha + 2e^+ + 2\nu_e
\end{equation}
The neutrino fluxes and spectra in these reactions are predicted by so called
Standard Solar Models (SSMs), e.g. BP2000\cite{ssm}.
The observed solar neutrino flux in all the experiments has significantly been
smaller than the expected.\cite{exp} From the recent results of solar neutrino
measurements, especially together with Super-Kamiokande and SNO, it comes from
neutrino oscillation in solar neutrinos. In order to determine the neutrino
oscillation parameter, mixing angle and delta mass square, it it important to
measure not only solar neutrino flux but also energy spectrum and flux time
variations, because those are independent of the uncertainties in the solar
models.

\subsection{Super Kamiokande}
 Super-Kamiokande is a 50000 tons of imaging water Cherenkov detector,
and it is located 1000m underground (2700m of water equivalent),
to shield against cosmic ray muons, in the Kamioka mine in Gifu Prefecture,
Japan. Cherenkov lights generated by charged particles scattered by neutrinos
in water are detected by 11146 20-inch photomultiplier tubes. The detector can
measure the energy spectrum of solar neutrinos very precisely because of the
well calibrated energy of recoil electrons.\cite{linac} And it can also measure
the time dependence of the solar neutrino flux, (day/night or seasonal
differences). The experiment started normal data taking from April 1st in 1996.
In this paper, the results of 1496 days of solar neutrino data,
which is from May in 1996 to July in 2001, are reported.
\section{Results of solar neutrinos in Super-Kamiokande}
\subsection{Flux}
 We have obtained 1496 days of solar neutrino data in the first phase of
Super-Kamiokande. The observed solar neutrino flux in Super-Kamiokande,
whose energy threshold is 5.0MeV, is
\begin{equation}
  2.35 \pm 0.02 (stat.) \pm 0.08 (sys.) \quad [\times 10^6/cm^2/sec].
\end{equation}
Comparing the result to standard solar model.
\begin{equation}
  \frac{Data}{SSM} = 0.465 \pm 0.005 (stat.) ^{+0.016}_{-0.015} (sys.),
\end{equation}
and the observed flux is significantly smaller than predicted flux.
And comparing with SNO charged current results\cite{sno}, the flux of
$\nu_{\mu,\tau}$ by neutrino flavor oscillation from $\nu_e$ should be
$3.45^{+0.65}_{-0.62}$.
\subsection{Seasonal variation}
\begin{figure}[p]
  \begin{center}
   \includegraphics[height=1.5in]{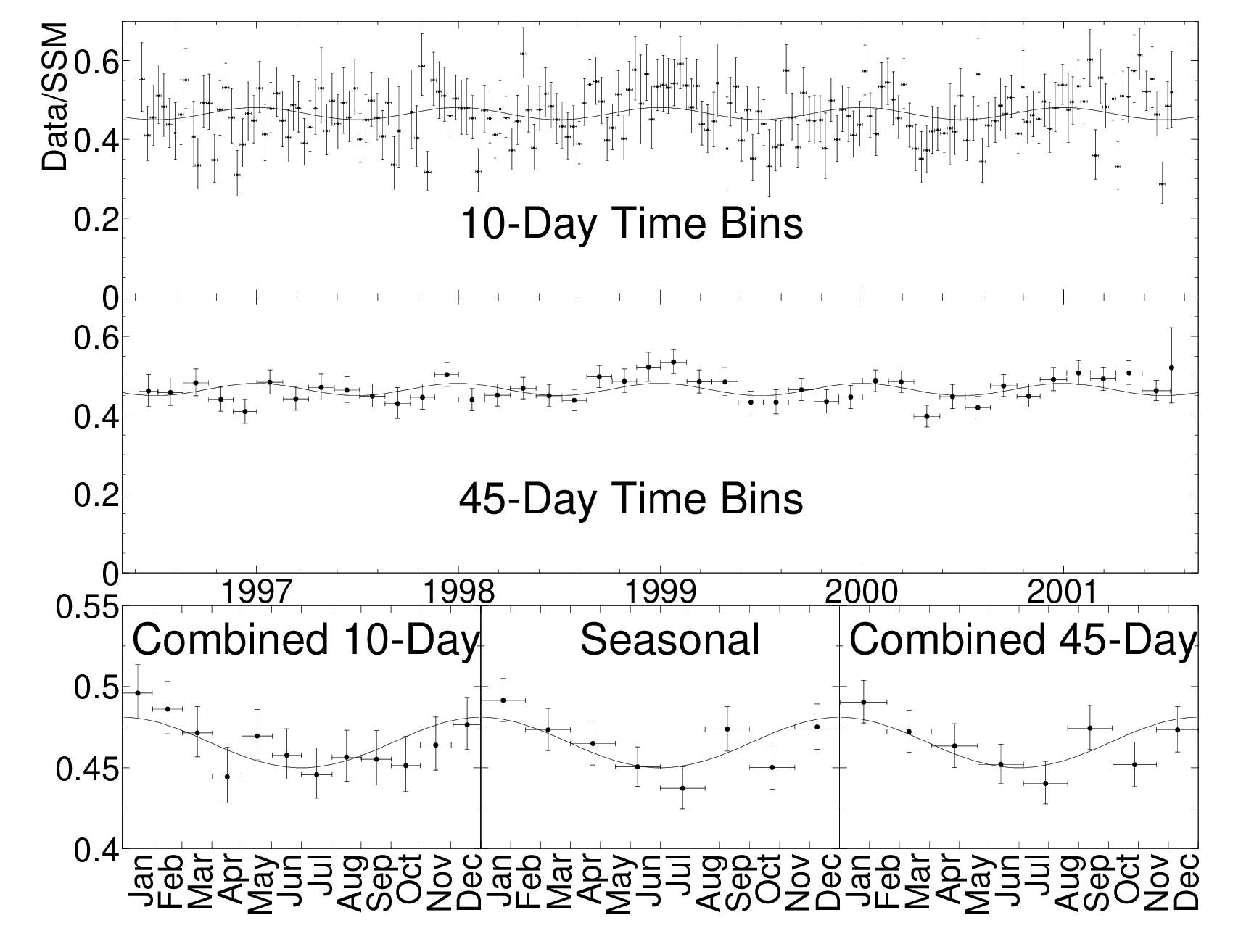}
  \end{center}
  \caption{Time variation of the solar neutrino flux in each time binning}
  \label{fig:time}
\end{figure}

 Fig~\ref{fig:time} shows the time variation of the solar neutrino flux,
the observed variation is consistent with the expected line.
No significant time variation of the solar neutrino flux can be seen
except for expected yearly variation by the eccentricity.
\subsection{Day/Night flux variation}
\begin{figure}[p]
  \begin{center}
   \includegraphics[height=1.5in]{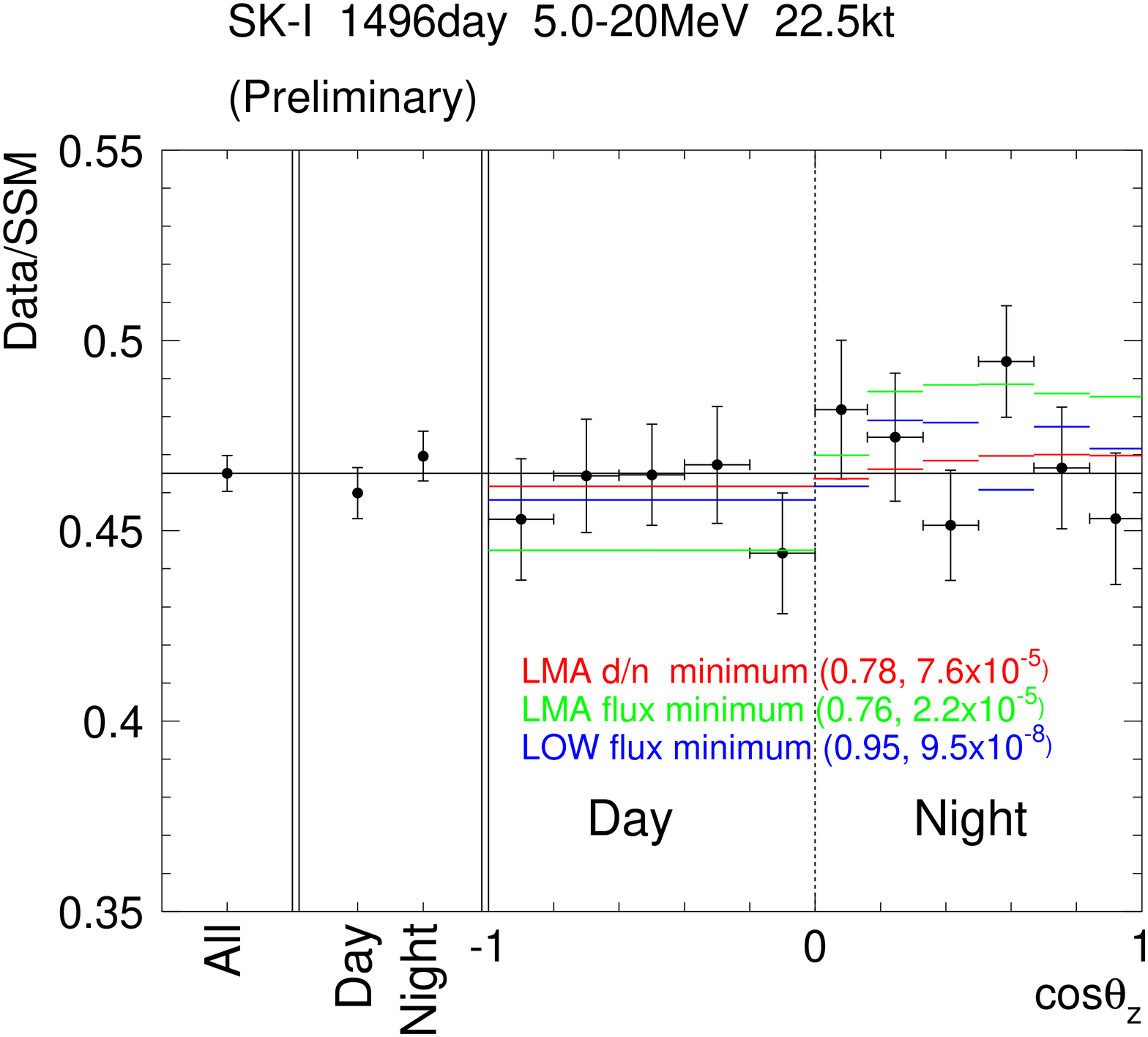}
   \includegraphics[height=1.5in]{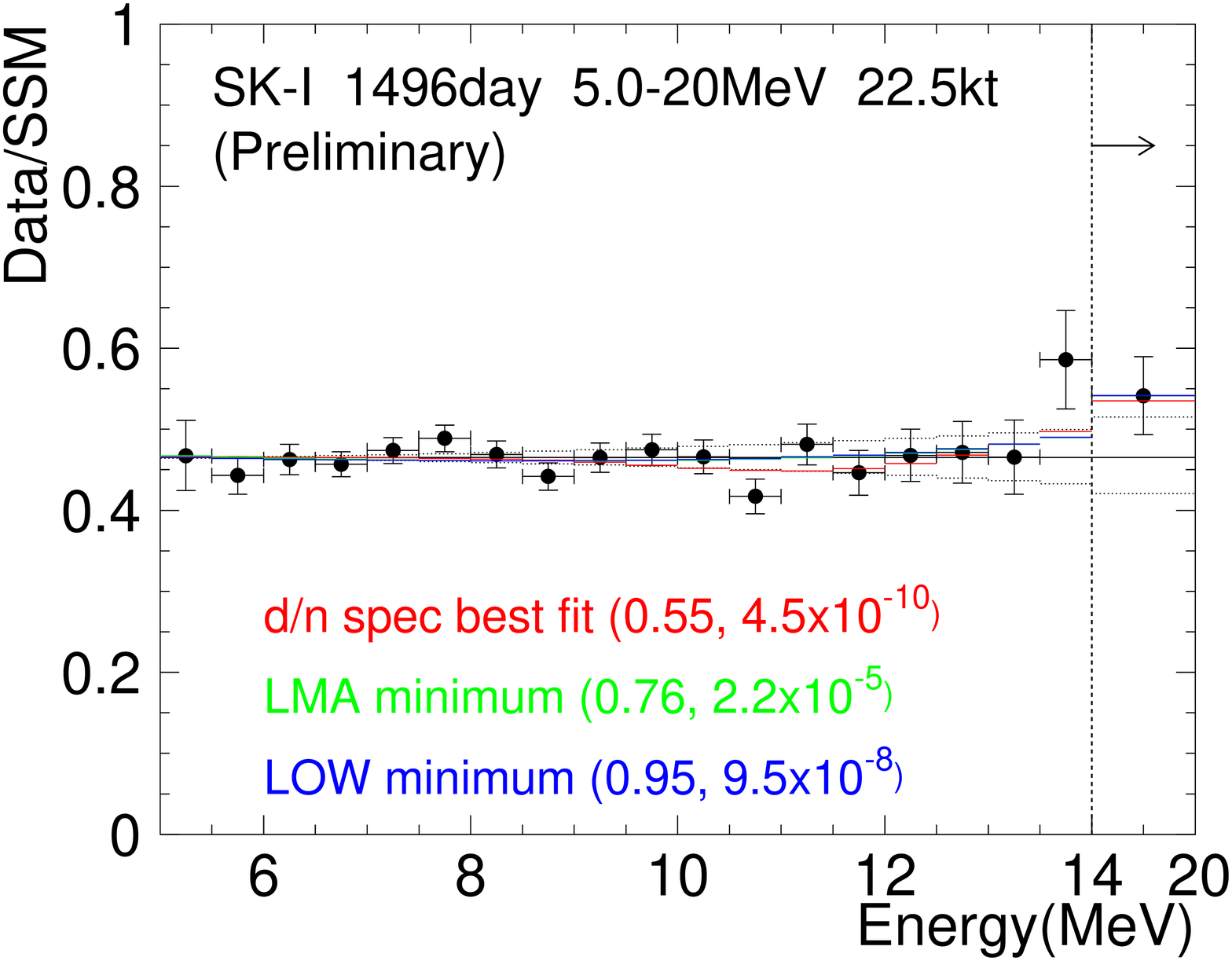}
  \end{center}
  \caption{Flux diveded by daytime and six night time bins (left).
           Energy spectrum of solar neutrino events normalized
           by the predicted energy spectrum (right).
           The expected flux assuming some neutrino oscillation parameters are
           also shown.}
  \label{fig:dnspe}
\end{figure}

 The day-night flux differences are also observed, and the result is,
\begin{equation}
  \frac{\phi_{day}-\phi_{night}}{(\phi_{day}+\phi_{night})/2}
 = -0.021 \pm 0.020 (stat.)  ^{+0.013}_{-0.012} (sys.),
\end{equation}
Fig~\ref{fig:dnspe} shows the solar neutrino flux in the daytime and
the night-time bins. The MSW effect\cite{msw} through the earth,
which is $\nu_e$ regeneration, could cause flux difference between daytime and
night-time. The flux ratio depends on distances through the earth and the
different electron density. For example, Fig~\ref{fig:dnspe} also shows the
expected solar neutrino flux in each bin assuming typical MSW parameters.
The plot is consistent with the flat, but also consistent with some
MSW solar neutrino oscillation parameters, e.g. LMA solution.
It can be used for the neutrino oscillation analysis as described
in the next section precisely.

\subsection{Energy spectrum}
 Fig~\ref{fig:dnspe} shows the observed spectrum of solar neutrino events
normalized by the predicted energy spectrum. We cannot see any distortion.
In this figure, the spectrum assuming the several couples of
neutrino oscillation parameters are also shown. It is also be used for
solar neutrino oscillation analysis.
\section{Solar neutrino oscillation analysis}
\subsection{Only Super-Kamiokande data}
\begin{figure}[p]
  \begin{center}
    \includegraphics[height=1.5in]{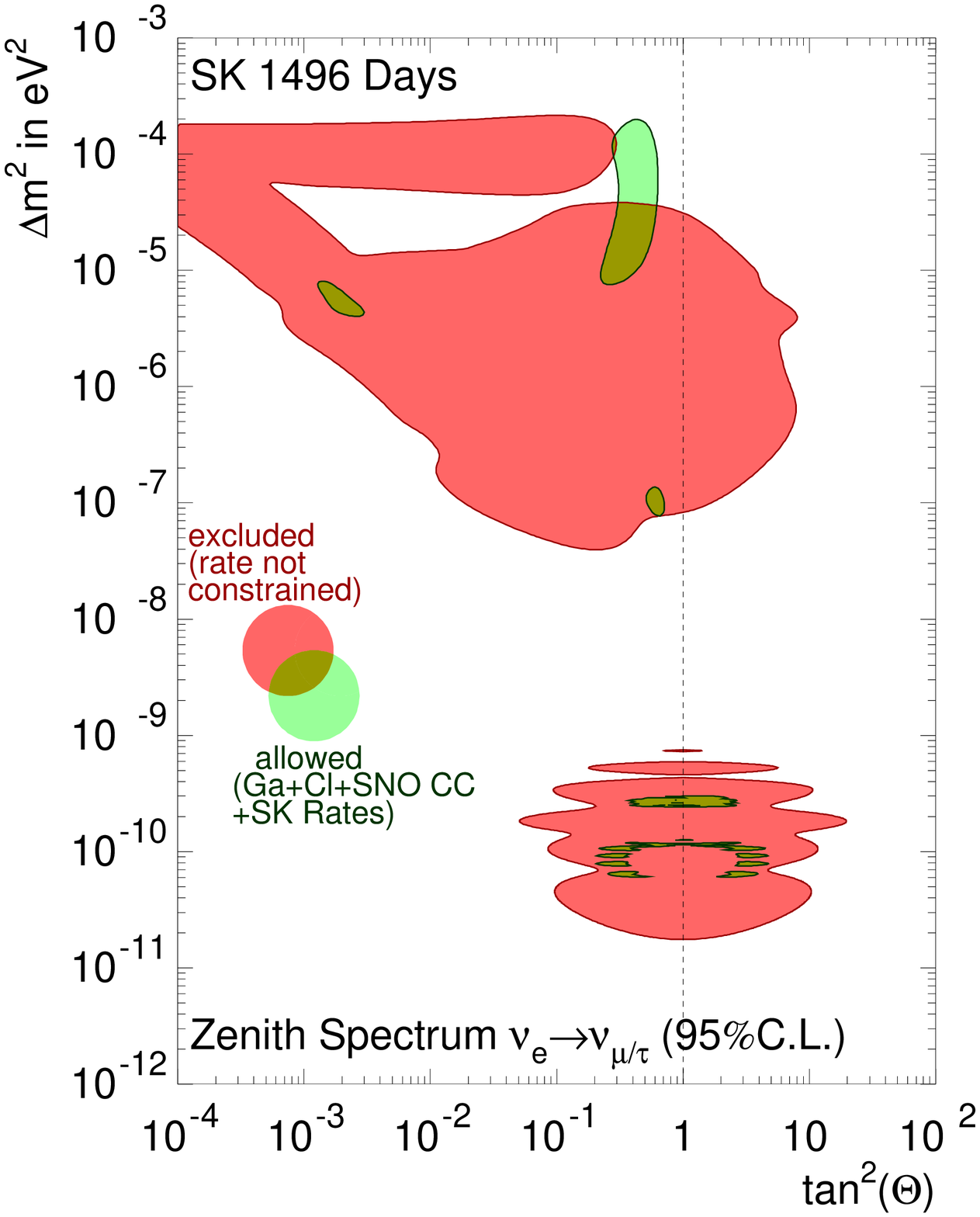}
    \includegraphics[height=1.5in]{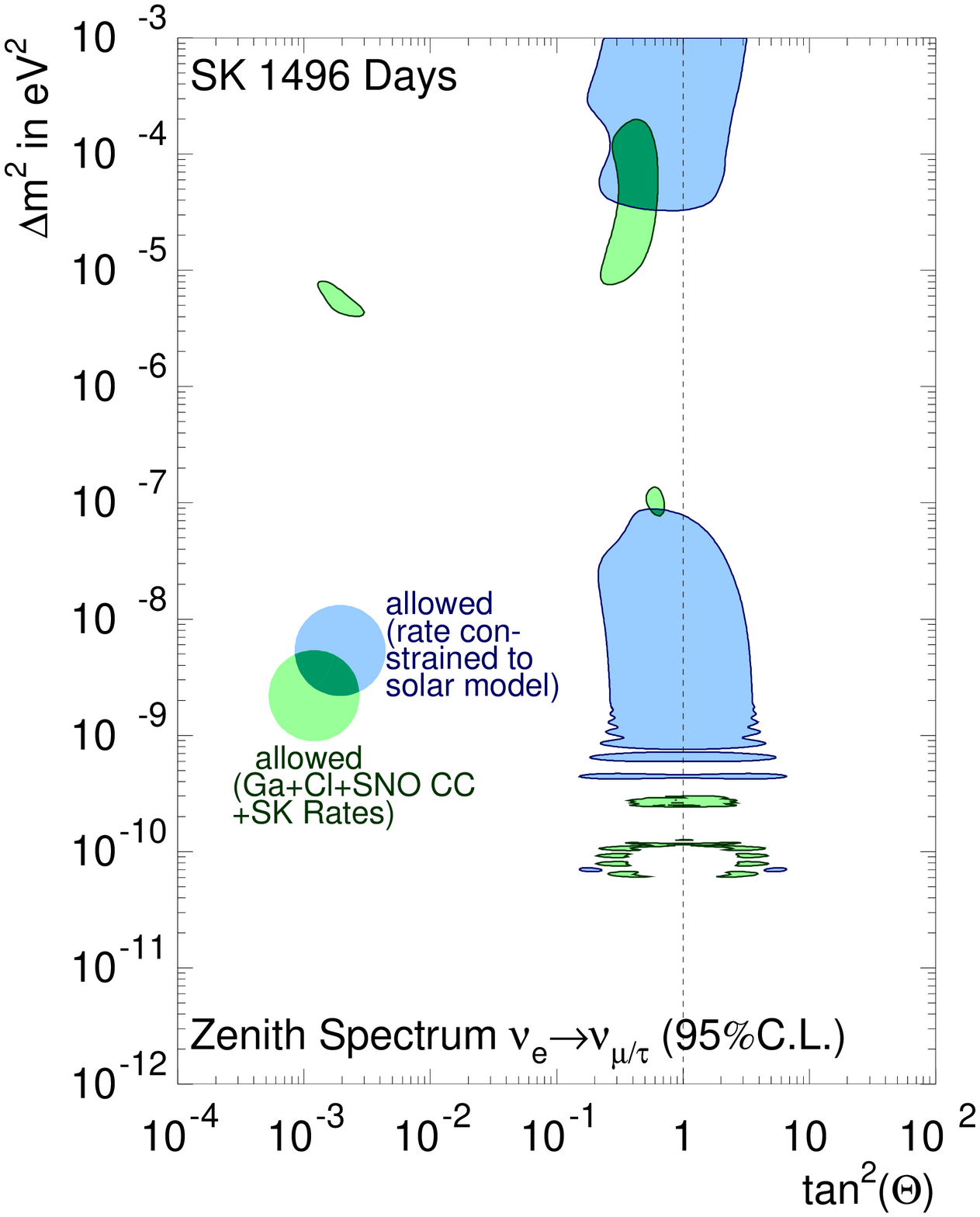}
    \includegraphics[height=1.5in]{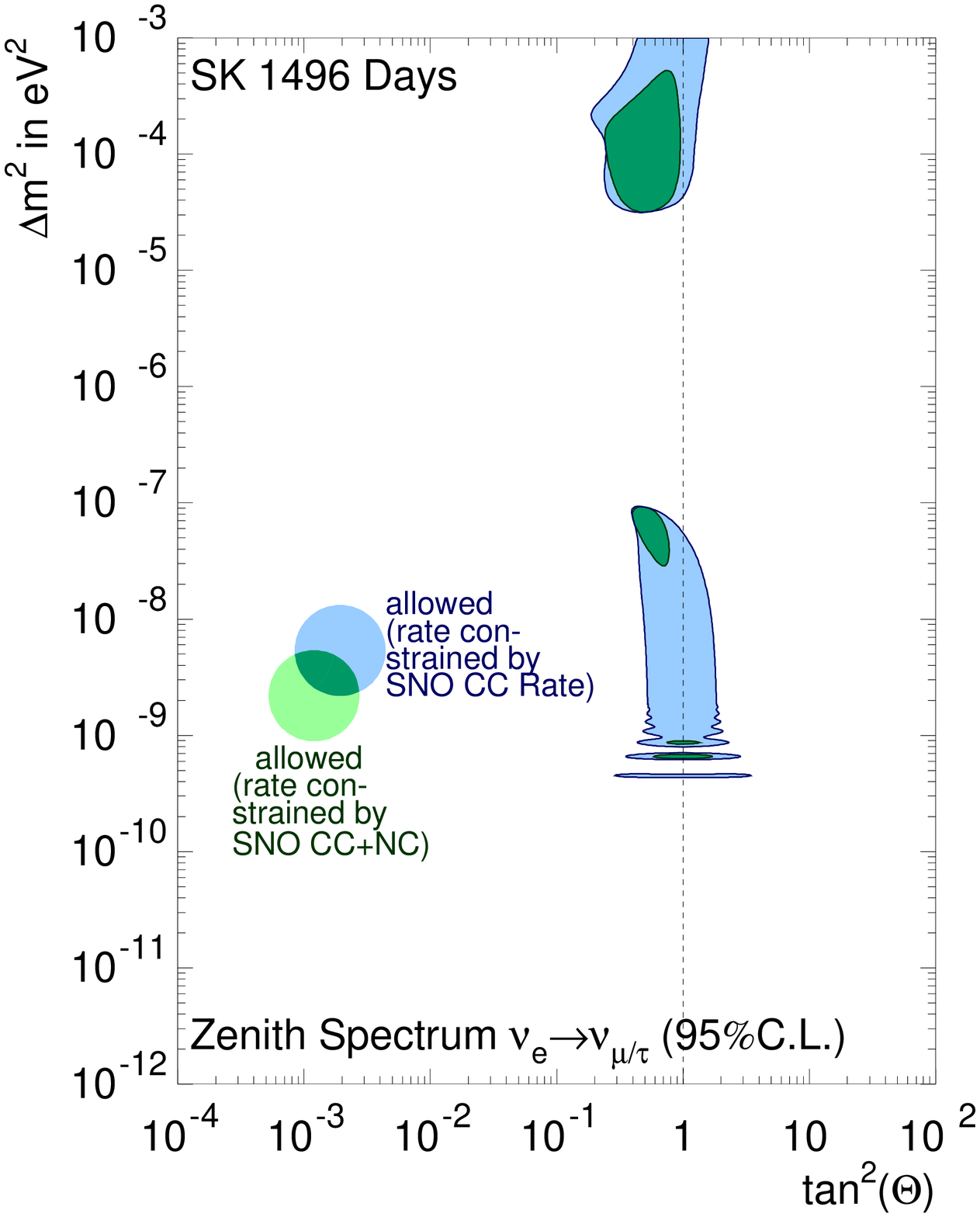}
    \includegraphics[height=1.5in]{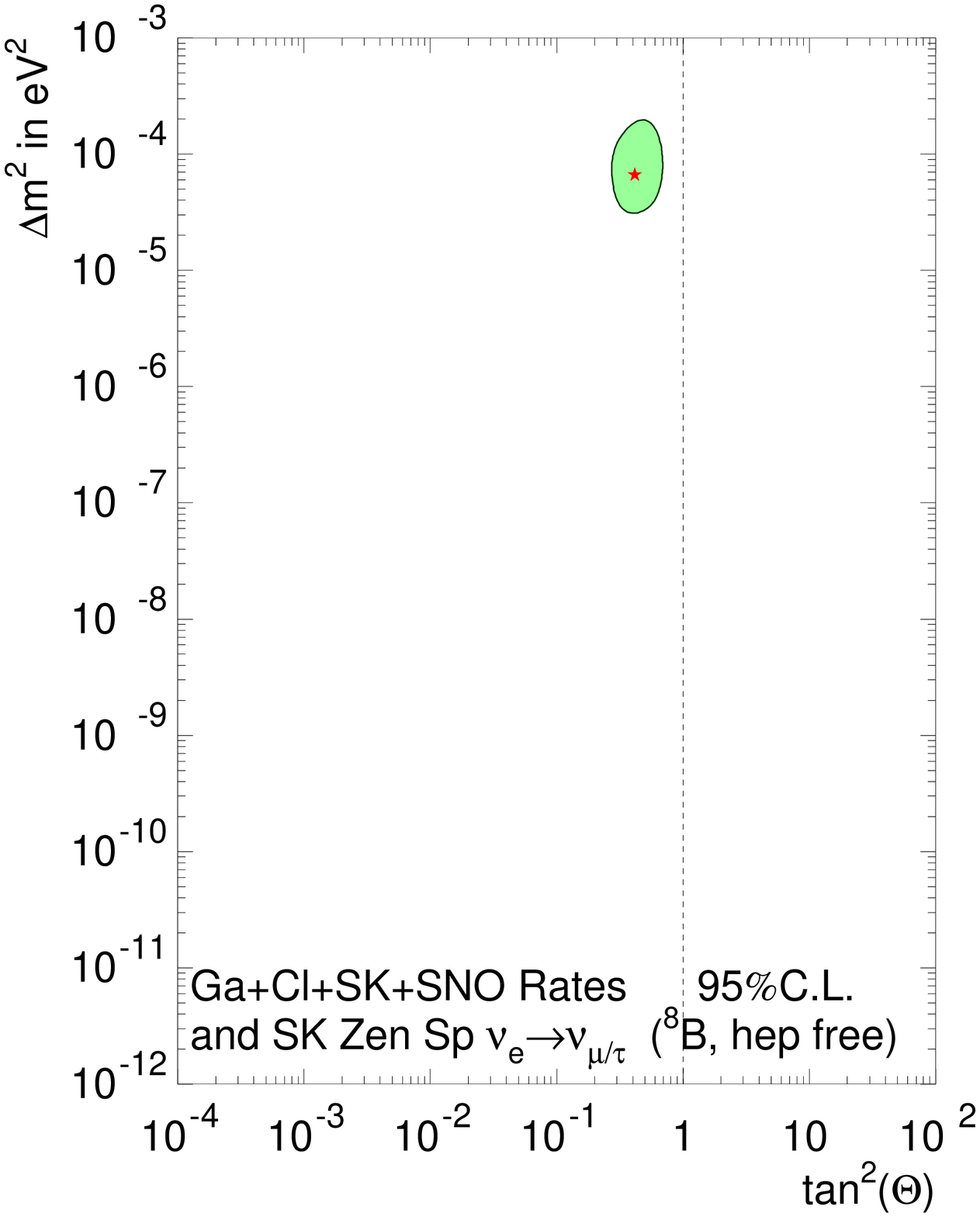}
  \end{center}
  \caption{The excluded region of neutrino oscillation parameter obtained
           from Zenith spectrum in Super-Kamiokande, which is overlaied with
           the allowed region obtained by the only flux results in all the
           solar neutrino. The 1st figure shows flux free and the 2nd shows
           flux constrainted analysis. The allowed region 
           from SK and SNO data with only SNO charged current (CC) data,
           and both CC and neutral current (NC) data. (3rd)
           The allowed region obtained by all the solar neutrino measurements.
           (4th)}
  \label{fig:osc}
\end{figure}

 For the neutrino oscillation analysis, we divide the Super-Kamiokande solar
neutrino data to day and six night bins in each six energy regions,
called Zenith spectrum. First, the left figure of Fig~\ref{fig:osc} shows the
exclude region from Zenith spectrum without flux constraint, which is
overlaied with the allowed region obtained by the only flux results in all the
solar neutrino experiment (green area). It can exclude some neutrino
oscillation solution, ``SMA'', ``LOW'' and vacuum solution at 95\% C.L.,
which are allowed from flux only results.
Next, when we add the flux constraint, the remaining allowed region is only
large mixing angle as shown in the right figure of Fig~\ref{fig:osc}
(blue area). From these analysis, Super-Kamiokande results
conclude that only the large mixing angle region can be allowed at 95\% C.L..
\subsection{Combined fits}
 Comparing Super Kamiokande and SNO, we can get the allowed region of
neutrino oscillation parametes. The left of Fig~\ref{fig:osc} shows the
95\% C.L. allwoed region. The important point of this analysis is free from
the uncertainty of solar model prediction. This analysis also conclude that
only the large mixing angle region is favored at 95\% C.L..

 Finally, we consider all the solar neutrino experimental data,
the remaining allowed region is only ``LMA'' small region. The right figure of
Fig~\ref{fig:osc} shows the region with 95\%C.L., and the confidence level
of solutions other than LMA are more than 98.1\%.
\section{Current status of the second phase of Super Kamiokande}
 The second phase of Super Kamiokande (SK2) has started since 10th of December,
2002. The main differences from the first phase are that the number of PMTs are
about half, and enclosed the acrilic and fiberglass cover on each PMT
for prevent the generation of shock even if PMT is broken.
The LINAC calibration in SK2 indicate the quality of the detector is like as
expected. And even in the rouph analysis, which is not apply spallation cut,
we can see the clear solar peak as shown in Fig~{fig:sk2peak}.
\section{Summary}
 Super-Kamiokande has measured precisely solar neutrino flux, recoil electron
spectrum and time variations of the flux. No significant time variation and
energy distortion appear from the 1496 days of data. The solar neutrino
oscillation analysis has been done using those data. The result from the
Super-Kamiokande Zenith spectrum data favor Large neutrino mixing at 95\% C.L.
And only LMA solutions remain at 98.1\%C.L. combined with flux results from
all solar neutrino experiments. The second phase of Super Kamiokande has
started, and the quality of the data is as expected.
\section*{Acknowledgments}
I gratefully acknowledge the cooperation of Kamioka Mining and
Smelting Company. My participation of this meeting was financially supported
by the Japanese Ministry of Education, Science and Culture.
Finally, I appreciate all the members who organized this meeting.

\begin{figure}[p]
  \begin{center}
    \includegraphics[height=1.5in]{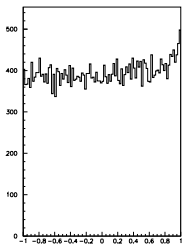}
  \end{center}
  \caption{The solar peak in SK2 data. The horizontal axis shows the angular
           between the solar direction and the SK2 events' direction.
           Solar direction is shown in value one of the horizontal axis.}
  \label{fig:sk2peak}
\end{figure}
\end{document}